\documentclass[twocolumn,showpacs,preprintnumbers,amsmath,amssymb]{revtex4}

\usepackage{graphicx}
\usepackage{dcolumn}
\usepackage{bm}
\usepackage{feynmp}
\usepackage{amssymb}

\newcommand{\be}{\begin{equation}}
\newcommand{\ee}{\end{equation}}
\newcommand{\bea}{\begin{eqnarray}}
\newcommand{\eea}{\end{eqnarray}}
\newcommand{\bean}{\begin{eqnarray*}}
\newcommand{\eean}{\end{eqnarray*}}
\newcommand{\bi}{\begin{itemize}}
\newcommand{\ei}{\end{itemize}}
\newcommand{\bdm}{\begin{displaymath}}
\newcommand{\edm}{\end{displaymath}}

\newcommand{\ie}{{\it i.e.}}

\newcommand{\lag}{{\mathcal L}}

\newcommand{\meg}{{\mu\to e\gamma}}

\newcommand{\mNeN}{{{\mu}N\to eN}}

\begin{document}

\preprint{ANL-HEP-CP-03-101, ALTA-TH-02-04}

\title{Differentiating solutions to the gauge hierarchy problem\\
through rare muon decays}

\author{David W.~Maybury}
\email{dmaybury@phys.ualberta.ca}
\affiliation{Department of Physics, University of Alberta, Edmonton AB T6G 2J1, CANADA.}

\author{Brandon Murakami}
\email{murakami@hep.anl.gov}
\affiliation{High Energy Physics Division, Argonne National Laboratory, Argonne, IL 60439 USA}

\date{\today}

\begin{abstract}
We demonstrate the potential of forthcoming $\meg$ and $\mu-e$ conversion experiments to implicate or disfavor solutions to the gauge hierarchy problem before the advent of the CERN Large Hadron Collider.  Solutions of dynamical electroweak symmetry breaking, little Higgs, supersymmetry, and extra dimensions are considered.  Correlations of $\meg$ and $\mu-e$ conversion branching ratios are analyzed for discriminating patterns.  Measurements of these exotic muon decays may have compelling implications for supersymmetric solutions.\\[1mm]

\noindent
\center
{\it A contribution to the 2004 Lake Louise Winter Institute proceedings.}
\end{abstract}


\maketitle

{\bf Motivation.}  What is the mechanism behind electroweak symmetry breaking (EWSB)?  Arbitrary extensions to the standard model (SM) generically exhibit charged lepton flavor violation (LFV).  If the forthcoming $\meg$ experiment MEG reports a discovery of LFV, what progress can be made if the vast majority of theoretical models claim it?  These are our motivating questions.  With the possible origins for LFV as wide as quantum field theory will allow, we are powerless to address arbitrary origins.  In this study, we restrict our focus to examining correlated signals of $\meg$ and $\mu-e$ conversion \cite{Kuno:1999jp} as a guide to implicating or disfavoring solutions to the gauge hierarchy problem (GHP), the quadratic divergence of the Higgs mass.

Perturbatively, the SM has an exact global symmetry for each generation of leptons.  Therefore, positive signal at MEG would be a clear indication of new physics.  Neutrino LFV is confirmed; thus one expects loop induced charged LFV, as the SM unifies the left handed charged and neutral leptons as an $SU(2)_L$ doublet.  However, the SM augmented by the seesaw yields unobservable LFV with rates suppressed by the high Majorana scale.

Solutions of the GHP are typically accompanied by incidental paths to observable LFV:  either through (1) effectively bringing the effects of neutrino oscillation phenomena down to a lower scale, such as with supersymmetric sleptons or extra dimensional Kaluza-Klein (KK) lepton states, or (2) providing new physics with tree-level LFV through exotic gauge or scalar bosons.  Here, we focus on the broad classes of dynamical EWSB, little Higgs, extra dimensions, and supersymmetry.  Each solution has many subclasses and many specific manifestations within.  An exhaustive analysis to attempt to reproduce the correlations of $\meg$ and muon conversion branching ratios is therefore a very demanding goal.  Therefore, we further restrict our scope and sketch general LFV origins of these solutions.

{\bf Strategy.}  For on-shell out-going photons, $\meg$ is inherently a dipole interaction and necessarily has a chirality flip.  The most general interactions may be written in Dirac notation as
\be
-\lag_{\meg} = em_\mu [\bar\mu \sigma^{\mu\nu}(A_2^RP_L + A_2^LP_R) e] F_{\mu\nu} + {\rm c.c.}
\ee

Muon conversion can be written in terms of scalar, pseudo-scalar, vector, pseudo-vector, and tensor interactions at the quark level:

\bea
-\lag_{\mu N \to eN}
&=& \sum_q \left\{e^2 Q_q [\bar\mu \gamma^\mu(A^L_1P_L + A^R_1P_R) e][\bar q\gamma_\mu q] \right.\nonumber\\
&& + e^2 Q_q [\bar\mu im_\mu\sigma^{\mu\nu}q_\nu (A^R_2P_L + A^L_2P_R) e][\bar q\gamma_\mu q] \nonumber\\
&& + [\bar e (a^L_{Sq}P_L + a^R_{Sq}P_R) \mu][\bar qq] \nonumber\\
&& + [\bar e (a^L_{Pq}P_L + a^R_{Pq}P_R) \mu][\bar q \gamma^5 q] \nonumber\\
&& + [\bar e \gamma^\mu(a^L_{Vq}P_L + a^R_{Vq}P_R) \mu][\bar q \gamma_\mu q] \nonumber\\
&& + [\bar e \gamma^\mu(a^L_{Aq}P_L + a^R_{Aq}P_R) \mu][\bar q \gamma_\mu\gamma^5 q] \nonumber\\
&& \left. + [\bar e \sigma^{\mu\nu}(a^L_{Tq}P_L + a^R_{Tq}P_R) \mu][\bar q \sigma_{\mu\nu} q] \right\}
\label{eq-mcoperators}
\eea
Here, $Q_q$ is the quark electric charge.  Although the first and fifth lines are redundant, for clarity, we separate the off-shell photon contributions from those from the $Z$ and exotic gauge bosons.  A complete calculation of the muon conversion rate requires encoding the quark level interactions into the nucleon level then to the nucleus level \cite{muonconversion}.

If muon conversion is dominated by on-shell photon exchange, then there will be a strict linear relationship between branching ratios of $\meg$ and muon conversion.  The correlated branching ratios differ roughly by an $\alpha$ and nuclear factor.  For example, in the case of aluminum \cite{Kuno:1999jp}, the model independent rate is
\be
{\rm BR}(\mu{\rm Al}\to e{\rm Al}) \approx {\rm BR}(\meg)/389 .
\label{eq-mcdictionary}
\ee

\begin{table}[t]
\center
\begin{tiny}
\begin{tabular}{ccccc}
Process & Leading experiment & BR reach & Future experiment & BR reach\\
\hline\hline
$\meg$ & MEGA (1999) & $1.2\times10^{-11}$\cite{Brooks:1999pu} & MEG (2005) & $4.5\times10^{-14}$\cite{Mori:sg}\\
\hline
$\mu N\to eN$ & SINDRUM II (1998) & $6.1\times10^{-13}$\cite{Wintz:rp} & MECO (2009) & $>10^{-16}$\cite{Hebert:kf} \\
&&& PRIME (2008) & $\sim10^{-18}$\cite{Kuno} \\
\hline
\end{tabular}
\end{tiny}
\caption{Past and near future LFV experiments.  The listed dates for the future experiments are the intended start dates of data acquisition.  PRIME may start as late as 2010.}
\label{tab-experiments}
\end{table}

Our motivating question then reduces to ``Which of the solutions of the GHP generically exhibit photon dominated muon conversion?''  The forthcoming LFV experiments are summarized in Table \ref{tab-experiments}.  The strategy then becomes to await a positive signal at MEG.  A positive signal at MEG virtually guarantees positive signals at MECO and PRIME, as they are able to probe $\meg$ further than MEG.  If the measured branching ratios are linearly correlated in the manner predicted by eq.~(\ref{eq-mcdictionary}), then we are provided a significant guide to differentiating GHP solutions.

It should be noted that while any model that exhibits LFV is capable of both $\meg$ and muon conversion, there is no reason {\it a priori} that muon conversion must be dominated by on-shell photon mediation of $\meg$.  Box diagrams or exotic gauge or scalar boson exchange may dominate, for example.  From this perspective, the question may be alternatively cast into asking which of the solutions require LFV bosons in their minimal incarnations.  Also, it should be emphasized that the monopole operators $A^L_1$ and $A^R_1$ are not present in BR($\meg$).  Therefore, photon mediation of muon conversion may still violate eq.~(\ref{eq-mcdictionary}) if the monopole operators are not neglible.

{\bf Dynamical EWSB.}
Technicolor is the prototype for dynamical EWSB models \cite{technicolor}.  However, pure technicolor only has the ability to generate masses for the weak bosons; fermions are not addressed.  This is accomplished through a gauge group ${\cal G}_{\rm TC}$ which becomes strongly coupled at a scale $\Lambda_{\rm TC}$ under which (fermionic) techniquarks form a fermion condensate.

To address the SM fermion masses, the gauge group is enlarged to the extended technicolor (ETC) gauge group ${\cal G}_{\rm ETC}$ which breaks to ${\cal G}_{\rm TC}$.  Through ETC gauge interactions, the SM fermions are coupled to the condensate or, in practical terms, a composite scalar vacuum expectation value (vev) which gives rise to spontaneously generated masses.  However, in practice, one finds conflict between acceptable masses and compliance with electroweak precision data.

Efforts to overcome all such problems has lead to a few classes of models (namely non-commuting ETC \cite{ncetc}, topflavor \cite{topflavor}, and top-color assisted technicolor \cite{topcolor}) that utilize non-universal gauge groups that include or break to a non-universal $U(1)$.  Upon rotation to the lepton mass state basis, the resulting $Z'$ will have LFV primal vertices.

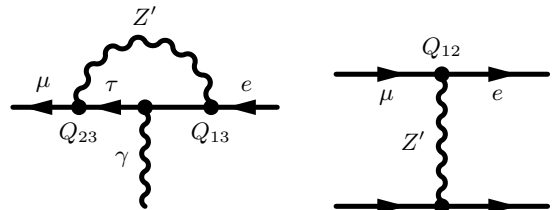
\begin{figure}[t]
\begin{center}
\begin{fmffile}{meg-zprime}
\begin{fmfgraph*}(100,75)
\fmfpen{thick}
\fmfleft{mu}
\fmfright{e}
\fmfbottom{ext}
\fmf{fermion,label=$e$}{e,v1}
\fmf{plain}{v1,v2}
\fmf{fermion,label=$\tau$}{v2,v3}
\fmf{fermion,label=$\mu$}{v3,mu}
\fmffreeze
\fmf{photon,right,label=$Z'$}{v1,v3}
\fmf{photon,label=$\gamma$}{v2,ext}
\fmfdot{v1,v2,v3}
\fmfv{label=$Q_{13}$,label.angle=-90}{v1}
\fmfv{label=$Q_{23}$,label.angle=-90}{v3}
\end{fmfgraph*}
\end{fmffile}
\begin{fmffile}{mc-zprime}
\begin{fmfgraph*}(100,50)
\fmfpen{thick}
\fmfleft{Nin,mu}
\fmfright{Nout,e}
\fmf{fermion,label=$\mu$}{mu,v1}
\fmf{fermion,label=$e$}{v1,e}
\fmf{fermion,label=$N$}{Nin,v2}
\fmf{fermion,label=$N$}{v2,Nout}
\fmffreeze
\fmf{photon,label=$Z'$}{v1,v2}
\fmfdot{v1,v2}
\fmfv{label=$Q_{12}$,label.angle=90}{v1}
\end{fmfgraph*}
\end{fmffile}
\end{center}
\caption{The dominant mass state amplitudes for $Z'$ induced $\meg$ and muon conversion.  For $\meg$, an internal tau propagator is dominant due to its mass insertion enhancement.}
\label{fig-zprime}
\end{figure}

Such LFV gauge bosons are capable of producing $\meg$ at loop level and muon conversion at tree level \cite{lfvzprimes}.  Tree level muon conversion will not have the dipole operator structure necessary to reproduce the linear correlation of eq.~(\ref{eq-mcdictionary}).  Take the example of a $Z'$ as in fig.~\ref{fig-zprime}.  In order for muon conversion to be on-shell photon dominated, the $Z'$ must have the combination $Q_{12} Q_{qq}$ either vanish or tuned sufficiently small simultaneously with $Q_{13} Q_{23}$ sufficiently large.  Whether or not a plausible model can supply these charges simultaneously avoiding excessive tau LFV is an open question.

{\bf Little Higgs.}
In general, there are two potential sources of LFV in little Higgs models \cite{littlehiggs}.  Flavor physics to generate the observed hierarchical fermion masses is required element that must exist in a complete theory of nature.  A previous study of quark flavor constraints \cite{Chivukula:2002ww} assumes the flavor physics is a gauged symmetry and  demonstrated, for composite Higgs models, such as little Higgs and dynamical EWSB models, the mass scale of the flavor-aware gauge bosons may be as low as roughly 100 TeV.  Via LFV $Z'$s, as with dynamical EWSB models, tree level muon conversion would be expected to dominate over the loop induced photon exchange.  No linear correlation with $\meg$ would result.

The second potential LFV source originates directly from concerns of EWSB naturalness.  The SM-like Higgs boson of little Higgs models descends from a larger representation of the model's global and local symmetries, such as $SU(5)/SO(5)$ for the littlest Higgs.  In general the decomposition of $\Sigma$ should result in exotic scalar representations that couple to ordinary matter, such as the littlest Higgs' complex triplet.  

Meanwhile, the large top mass induces a correction to the Higgs mass squared proportional to $\Lambda^2/16\pi^2$, ruining the naturalness solution.  Models to date address this by introducing exotic top-like vector pair $\tilde t$ and $\tilde t'$.  By design, along with some tuning, the symmetry structure provides a Higgs mass cancellation between top and top-like quarks.

As the top sector is the most relevant to EWSB fine tuning arguments, only the top sector is increased by the additional matter by current models.  However, if additional matter were included for all quark and lepton flavors, LFV would arise through the exotic scalar multiplets, \ie~the complex triplet of the littlest Higgs \cite{Han:2003wu}.  This results in tree level muon conversion through scalar exchange and no linear correlation with $\meg$.

{\bf Supersymmetry.}
Perturbatively, the minimal parameterizations of $R$-parity conserving gravity, gauge, and anomaly mediation do not have LFV at any scale.  However, there are three motivated extensions that may introduce LFV:  neutrino masses and oscillations \cite{neutrinolfv}, flavor dynamics that produce the fermion mass hierarchy \cite{flavorlfv}, and unified gauge theories (GUTs) \cite{gutlfv}.  Off-diagonal trilinear couplings and $R$-parity violation are also sources of LFV but less motivated and not considered here.  While GUTs may not be required in nature, flavor dynamics are.  Therefore, beyond a supersymmetric standard model (as with all extensions) there must lie a potential source for LFV.

Quite rigidly for the MSSM, muon conversion is mediately nearly exclusively through photon exchange, rendering irrelevant $Z$ penguins, Higgs penguins, and box diagrams involving neutralinos or charginos.  Central to our goals, it is necessary to understand why muon conversion is dominated by on-shell photon exchange.  There are four contributing factors:
\begin{enumerate}
\item The muon rest mass is the maximum available energy to a bound muonic atom in ground state.  Therefore, each sparticle or $Z$ propagator is at least an ${\cal O}(m_\mu/m_Z) \sim 10^{-3}$ suppression relative to the photon propagator.  However, this type of suppression is common to all heavy sources of LFV.
\item Photon penguins involving Higgsinos or left-right slepton mass mixing have $\tan^2\beta$ enhanced probability in the dipole operator couplings.
\item The smallness of the lepton Yukawa couplings greatly suppress Higgs exchange.  However, Higgs exchange may be relevant in the case of a light CP even heavy Higgs $H$ (under $\sim$ 400 GeV) and large $\tan\beta$ (over $\sim$ 60) \cite{Kitano:2003wn}.
\item All box diagrams necessarily involve squarks, at leading order.  Squarks are generically heavy in the MSSM due to the renormalization group (RG) flow, given boundary conditions at a high scale.  RG flow is determined by a model's particle content, which is fixed by definition of the MSSM.
\end{enumerate}
There are also several reasons due to the {\it lack} of certain elements in the MSSM, \ie~the lack of non-universal gauge bosons \cite{Murakami:2001hk}, additional multiplets that extend the lepton sector, etc.

Point 2 above is a direct consequence of the MSSM.  Supersymmetry requires a minimum of two Higgs doublets for the cancellation of chiral anomalies.  This, in turn, allows for the existence of $\tan\beta$.  If not for this property, the monopole operators may be of the same order as the dipole operators, preventing relation eq.~(\ref{eq-mcdictionary}).  Futhermore, the doublets must be of the Type II class due to the analyticity of the superpotential.  If not for this property, tree level LFV may arise from the Higgs sector.

Aside from the interesting but less probable Higgs mediation case, these reasons sum to the generic prediction of on-shell photon mediated muon conversion in the MSSM.  Due to the large color coupling $g_3$, point 4 above is very robust even with the addition of exotic matter, despite the supersymmetry breaking mechanism.  Therefore, one may venture to say supersymmetry with plausible supersymmetry breaking, quite generally, favors on-shell photon mediated muon conversion.  As the experimental implication of gauge coupling unification is the strongest indirect evidence for the MSSM to date, we emphasize linearly correlated $\meg$ and muon conversion rates would be another significant experimental implication.

{\bf Extra dimensions.}
Because other gauge hierarchy solutions are often set in extra dimensions (such as supersymmetry and dynamical EWSB), we limit this discussion to those classes in which the extra dimension directly solve the hierarchy problem.  The remaining two classes are those dilute high scales through volume suppression in flat space and those that dilute scales through a gravitational warp factor.

In itself, extra dimensions say nothing about flavor.  However, motivated uses that involve flavor are those that address the neutrino oscillation puzzle and the fermion mass hierarchy.  Both usages generally lead to LFV.  Right handed neutrino singlets in the bulk, warped or flat, will generate $\meg$ at one loop where Kaluza-Klein (KK) states play a similar role as supersymmetric sparticles \cite{Kitano:2000wr}.

However, unlike the MSSM case, there is no $\tan\beta$ enhancement of the dipole operators over the monopole operators.  Therefore, the linear correlation of eq.~(\ref{eq-mcdictionary}) does not hold, despite the dominance of photon mediation in muon conversion.  While it is possible to create $\tan\beta$ enhancement via the introduction of Type II Higgs doublets in these scenarios, it would not be as motivated as supersymmetry's simple requirement for it.

The fermion masses may be understood as their fields being localized in the bulk where cartography determines the Yukawa couplings through overlap with the Higgs field \cite{splitfermions}.  LFV studies in such scenarios have been performed in flat \cite{Barenboim:2001wy} and warped \cite{warpedcartography} scenarios.  For flat space, the Higgs may propagate throughout the bulk; a fermion's mass is proportional to the separation of its left and right chiral states.  The zero modes of the photon and $Z$ fields are constant over the extra dimensions, but the KK states vary over the extra dimensions.  For warped space, the Higgs is localized around the TeV brane; a fermion's mass is proportional to its overlap with the Higgs.  Only the zero mode of massless fields (the photon) is constant, while the zero mode of the $Z$ varies.  Gauge fields that vary over the extra dimension will have different overlap with localized fermions and thus, non-universal couplings, thus leading to FCNC.  Therefore, flat or warped, only massive gauge bosons mediate muon conversion at tree level in these scenarios.  No linear correlation results.

{\bf Outlook.}
The nearest next generation LFV experiment is MEG of the Paul Scherrer Institut (PSI), scheduled for data acquisition in 2005.  Should MEG see a positive signal of $\meg$, most theoretical models can claim it.  Future muon conversion experiments MECO and PRIME (see table 1) are then guaranteed a positive signal.  We emphasize that these LFV experiments are positioned to possibly conclude data acquisition and analysis before the LHC acquires meaningful statistics.

While there are many origins for LFV, we focus only on those that LFV origins that arise through motivated solutions of the GHP.  Through correlations of the observed branching ratios, it may be possible to disfavor large classes of solutions.  Measurements of BR($\meg$) and BR($\mNeN$) in the linear manner of eq.~(\ref{eq-mcdictionary}) would have compelling implications for the MSSM.  Non-compliance of eq.~(\ref{eq-mcdictionary}) is less illuminating, as it points to many possibilities described above, including solutions of dynamical EWSB, little Higgs, and extra dimensional scenarios.

{\bf Acknowledgments}
We thank B.~Campbell, E.~Jankowski, T.~Tait, and C.~Wagner for helpful comments.  We are grateful to R.~Kitano for pointing out the relevance of the monopole operators in extra dimensional scenarios.  B.M.~is supported by the U.S.~Department of Energy under contract W-31-109-Eng-38.  D.M.~is supported by the Natural Sciences and Engineering Research Council of Canada.

\end{document}